\documentclass[12pt,letterpaper]{article}          
\usepackage{osajnl}
\usepackage{amssymb}

\begin{document}

\title{Non-magnetic nano-composites for optical and infrared negative refraction index media}

\author{Robyn Wangberg$^1$, Justin Elser$^1$, Evgenii E. Narimanov$^2$, 
\\ and Viktor A. Podolskiy$^{1,*}$}

\address{$^1$ Physics Department, Oregon State University,
\\ 301 Weniger Hall, Corvallis OR 97331
\\$^2$ EE Department, Princeton University, Princeton NJ 08540}

\email{$^*$vpodolsk@physics.orst.edu}

\begin{abstract}
We develop an approach to use nanostructured plasmonic materials as a non-magnetic negative-refractive index system at optical and near-infrared frequencies. In contrast to conventional negative refraction materials, our design does not require periodicity and thus is highly tolerant to fabrication defects. Moreover, since the proposed materials are intrinsically non-magnetic, their performance is not limited to proximity of a resonance so that the resulting structure has relatively low loss. We develop the analytical description of the relevant electromagnetic phenomena and justify our analytic results via numerical solutions of Maxwell equations. 
\end{abstract}


\maketitle 

\section{Introduction}
The electromagnetic response of negative refractive index materials (NIM) \cite{veselago,PhysTodayLHM,pendry,shvetsPRB,efros,agranovich,podolskiyPRB,kivshar} has recently attracted unprecedented attention. Novel optical phenomena, predicted to take place in these unique systems include reversal of Snell Law, Doppler Effect, Cherenkov Effect\cite{veselago}, aberration-free \cite{veselago,efros,PhysTodayLHM} and sub-diffraction\cite{pendry,zhangSuperlens,podolskiyResolut,smithRESOLUT,merlin,webbSuperlens,smolyaninov,eleftheriades04,shvetsSuperlens} imaging, and excitation of the new types of surface and nonlinear waves \cite{kivshar,zakhidov}. In particular, realization of NIMs may potentially lead to fabrication of new types of lenses and prisms\cite{veselago,efros,pendry}, new lithographic techniques\cite{zhangSuperlens,shvetsSuperlens}, novel radars, sensors, and telecommunication systems. However, despite the great advantages NIM has to offer for optical and infrared spectral range, all practical realizations of NIM are currently limited to GHz frequencies\cite{smith,Parazzoli,sridhar,optExpLu}. 

Until recently there were two major approaches for NIM design. The first one is based on the original proposal\cite{veselago} that material with simultaneously negative dielectric permittivity and magnetic permeability must have a negative refraction index. This particular approach also benefits from the possibility to resonantly excite the plasmon polariton waves at the interface between NIM and surrounding media, which in turn may lead to sub-diffraction imaging\cite{pendry,zhangSuperlens,podolskiyResolut,smithRESOLUT,merlin,webbSuperlens,smolyaninovSuperlens}. However, the absence of natural magnetism at high (optical or infrared) frequencies\cite{landauECM} requires the design and fabrication of nanostructured meta-materials, to achieve the non-trivial magnetic permeability\cite{Parazzoli,smithTHz,shalaevWires,pendryNL,JNOPM2002,podolskiyOptExp,soukoulisTHz}. As these engineered systems typically operate in close proximity to resonance, resonant losses become the dominant factor in system response, severely limiting the practicality of resonant-based systems\cite{merlin,webbSuperlens,smithRESOLUT,podolskiyResolut,podolskiyResolut2}. 

The second approach for NIM design involves the use of photonic crystals\cite{sridhar,notomiPC,efrosPhotCryst,optExpLu,shvetsPRB}. However, the NIM response in these systems is typically associated with second or other higher-order bands and requires a complete bandgap between the band in use and the next band. The dispersion and very existence of the required bandgap are typically strongly affected by crystal disorder, unavoidable during the fabrication step. The manufacturing of the optical photonic crystals-based NIM typically requires 3D patterning with 10-$nm$ -- accuracy, which is beyond the capabilities of modern technology. 

To address the mentioned-above shortcomings of the traditional NIM schemes, we have recently introduced an alternative approach to design the NIM structure\cite{podolskiyPRB}. In contrast to ``conventional'' systems, the proposed design does not rely on either magnetism or periodicity to achieve negative refraction response. It has been shown that the combination of strong anisotropy of the dielectric constant and planar waveguide geometry yields the required negative phase velocity in the system\cite{podolskiyPRB}. Here we present the detailed description of NIMs proposed in Refs.~[\onlinecite{podolskiyPRB,podolskiyJMO}], study the effects related to waveguide boundaries, important for optical domain, and suggest several nanostructured materials providing the low-loss negative refraction response at optical and infrared frequencies. 

The rest of the paper is organized as follows: the next Section is devoted to EM wave propagation in strongly anisotropic waveguides; Section 3 describes the proposed realizations of the structure; imagining properties of these composites are shown in Section 4; Section 5 concludes the paper. 

\section{Negative refraction in strongly anisotropic waveguides}
\begin{figure}[htbp]
\centering
\includegraphics[width=8.3cm]{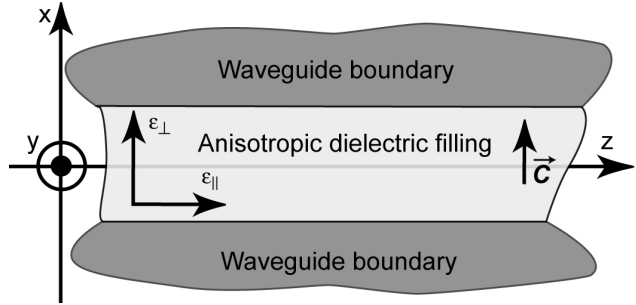}
\caption{Schematic configuration of non-magnetic negative-refraction system}
\label{figConfig}
\end{figure}

We consider wave propagation in the $2D$ planar waveguide structure shown in Fig.~\ref{figConfig}. The propagation in the system is allowed in the $y$ and $z$ directions, while the waveguide walls occupy the regions $|x|>d/2$. The waveguide core is assumed to be a homogeneous, non-magnetic ($\mu=1$) material, with a uniaxial anisotropic dielectric constant with dielectric permittivities $\epsilon_{\perp}$ and $\epsilon_{\|}$ along and perpendicular to the optical axis respectively. The optical axis of the core material ($C$) is assumed to be perpendicular to the direction of the wave propagation in the media ($C\|x$). Therefore, despite the anisotropy of the system, the effective refractive index of propagating in the planar geometry waves will be completely isotropic. 

Any wave propagating in the system can be represented as a linear combination of the {\it waveguide modes}\cite{podolskiyPRB,landauECM}. An individual mode is defined by its structure along the optical axis direction ($C$) and its polarization. Two different kinds of modes have to be distinguished. The modes of the first kind (known as TE waves) have their $E$ vector perpendicular to the optical axis. The propagation of these waves is fully described by the in-plane dielectric constant $\epsilon_{\|}$. The modes of the second kind (known as TM waves) have their $H$ vector in the waveguide plane and are affected by {\it both} $\epsilon_{\perp}$ and $\epsilon_{\|}$. The existence of these TM waves is crucial for the NIM described here.

In the analytical results presented below we limit ourselves to the case of single-mode propagation. We note that such a description provides complete information about the linear properties of the waveguide structure. Indeed, as mentioned above, an arbitrary wavepacket in the system can be represented as a linear combination of modes. In our numerical simulations discussed in Section 4 we utilize this property to compute the imaging performance of the system. 

\subsection{Waveguide with perfectly conducting walls}
As it has been shown in Ref.~[\onlinecite{podolskiyPRB}], the propagation of a mode in a planar waveguide can be described by the free-space-like dispersion relation: 
\begin{equation}
\label{eqDispEq}
k_y^2+k_z^2=\epsilon\nu{\rm k}^2,
\end{equation}
where $\epsilon$ is $\epsilon_{\|}$ for TE modes and $\epsilon_{\perp}$ for TM ones, $k_y$ and $k_z$ are the propagation components of the wavevector, ${\rm k}=\omega/c$ with $\omega$ and $c$ being the free-space angular frequency of the radiation, and speed of light in a vacuum; the propagation constant $\nu$ is given by
\begin{equation}
\label{eqNu}
\nu=1-\frac{\kappa^2}{\epsilon_{\|}{\rm k}^2},
\end{equation}
and the parameter $\kappa$ defines the mode structure in $x$ direction. 

As it directly follows from Eq.~(\ref{eqDispEq}), the phase velocity of a propagating mode is equal to 
\begin{equation}
\label{eqPhase}
v_p=n \; {\rm k},
\end{equation}
where the effective refraction index $n^2=\epsilon\nu$. Note that similar to the case of the plane wave propagation in free-space, the refraction index contains a product of two (mode-specific) {\it scalar} constants. A transparent structure must have both propagation constants of the same sign. The case of positive $\epsilon$ and $\nu$ corresponds to ``conventional'' (positive refraction index) material. The case of negative $\epsilon$ and $\nu$ describes NIM\cite{podolskiyPRB,podolskiyJMO}. The NIM behavior can be easily illustrated by comparing the Poynting vector $S_z$ and the wavevector $k_z$ as shown below. 

Similar to any waveguide structure, the mode in the system described here can be related to the $x$ profile of the longitudinal field component (the detailed description of such a dependence is given in Ref.~[\onlinecite{podolskiyPRB}]). To better illustrate the physical picture behind the mode propagation, in this section we present the results for the important case of perfectly conducting waveguide walls. In this case, the EM energy is confined to the waveguide core and the longitudinal field has a $\cos(\kappa x)$ or $\sin(\kappa x)$ profile depending on the symmetry with respect to the $x=0$ plane, with $\kappa=(2 j+1)\pi/d$ for symmetric and $\kappa=2\pi j/d$ for anti-symmetric modes respectively with $j$ being the integer mode number. The deviation from this idealized picture due to finite conductance of the waveguide material does not change the physical picture described in this section, and for the practical case of ``good'' metals (Ag,Al,Au) at near-IR to THz frequencies can be treated perturbatively. Results of such a perturbation approach are presented in the Section 2b.

The electric ($U_E$) and magnetic ($U_H$) field contribution to the energy density of a mode in weakly-dispersive material ($|\epsilon/\omega|\gg |{\rm d}\epsilon/{\rm d}\omega|$) is given by $U_E=\frac{1}{8 \pi d}\int (\mathbf{D}\cdot \mathbf{E^*}) dx$ and $U_H=\frac{1}{8\pi d}\int (\mathbf{H}\cdot \mathbf{H^*}) dx$ respectively\cite{landauECM} (the asterisk ($^*$) denotes the complex conjugation). Using the explicit mode structure for TE and TM waves (see Ref.~[\onlinecite{podolskiyPRB}]) we arrive to: 
\begin{eqnarray}
\label{eqEn}
U^{(TM)}_E&=&U^{(TM)}_H=\frac{1}{16\pi}\frac{\epsilon_{\|}^2 {\rm k}^2}{\kappa^2}|A_0|^2;\; 
U^{(TM)}=U^{(TM)}_E+U^{(TM)}_H=\frac{\epsilon_{\|}^2 {\rm k}^2}{8\pi\kappa^2 
}|A_0|^2; 
\\
U^{(TE)}_E&=&U^{(TE)}_H=\frac{\epsilon_{\|}}{16\pi}|A_0|^2; \;
U^{(TE)}=U^{(TE)}_E+U^{(TE)}_H=\frac{\epsilon_{\|}}{8\pi}|A_0|^2, 
\end{eqnarray}
where $A_0$ is the mode amplitude. Thus, extending the similarity between the waveguide system described here and the free-space propagation, the EM energy of any propagating wave is always positive and contains equal contributions from the electric and magnetic components of the field. 

It is also seen that the TE mode is in some sense very similar to the conventional plane wave propagating in the isotropic homogeneous dielectric. Namely, (i) energy density of the TE waves is exactly equal to that of the plane waves; (ii) there is no wave propagation in material with $\epsilon_{\|}<0$. In contrast to this behavior, the sign of the dielectric permittivity alone does not impose limitations on the propagation of TM modes. 

Another important characteristic of the energy transport in the EM system is the average energy flux given by the propagating component of the Poynting vector $\mathbf{S}=\frac{c}{4\pi}[\mathbf{E}\times \mathbf{H}]$. Selecting the direction of the wave propagation as $z$ axis, we obtain: 
\begin{eqnarray}
\label{eqPoynting}
S^{(\{TE;\;TM\})}_z=c\frac{k_z}
{\epsilon_{\{\|;\;\perp\}} {\rm k}}U^{(\{TE;\;TM\})}
\end{eqnarray}

It is clearly seen from Eq.~\ref{eqPoynting} that the relation between the direction of the phase velocity and energy flux is defined by the sign of the dielectric constant (for a given mode polarization) -- positive $\epsilon$ means $n>0$ propagation, while $\epsilon<0$ signifies the NIM case. Of course, for this relation to take place, we must require the medium to be transparent -- both propagation constants $\epsilon$ and $\nu$ must be of the same sign. As it can be seen from Eq.~(\ref{eqDispEq}), the NIM condition can be satisfied only for TM wave and only in the case of extreme anisotropy of the dielectric constant of the core material ($\epsilon_{\|} \epsilon_{\perp}<0$). The feasibility of the fabrication of such unusual materials will be discussed in the Section 3. 

\subsection{The effect of finite wall conductance}
In this section we consider the practical realization of the system described above, in which the anisotropic core material is surrounded by metallic walls. The electromagnetic properties of metals at high (GHz to optical) frequencies are dominated by the dynamics of the free-electron plasma-like gas. Following the approach described in e.g. Ref.~[\onlinecite{landauPK}] it is possible to write down the high-frequency effective permittivity of metal in Drude form: 
\begin{equation}
\label{eqDrude}
\epsilon_m(\omega)=\epsilon_{\infty}-\frac{\Omega_{\rm pl}^2}{\omega(\omega+i \;\tau)},
\end{equation}
where the constant term $\epsilon_{\infty}$ describes the contribution of the bound electrons, $\tau$ is responsible for EM losses due to (inelastic) processes, and $\Omega_{\rm pl}=\frac{N_e\;e^2}{m_{\rm eff}}$ is the plasma frequency with $N_e, e$, and $m_{\rm eff}$ being the free-electron concentration, charge, and effective mass respectively. Note that for $\omega<\Omega_{\rm pl}/\sqrt{\epsilon_{\infty}}$ the permittivity of the metal becomes negative $\epsilon^{\prime}_m<0$ (here and below single and double prime ($^{\prime};\;^{\prime\prime}$) denote the real and imaginary parts respectively). For most of ``good'' metals (Ag,Al,Au) the plasma frequency is of the order of $10\;eV$ and $\epsilon_{\infty}\approx 1$, which means that $\epsilon_m^{\prime}$ is negative from optical to GHz frequencies. The losses, given by the parameter $\epsilon_m^{\prime\prime}/|\epsilon_m^{\prime}|\ll 1$ are typically small in these spectral ranges. 

Similar to the case of perfectly conducting waveguide walls, the structure of the modes in the system can be still derived from the dependence of the longitudinal ($z$) field component on the $x$ coordinate, which has $\cos(\kappa x)$ or $\sin(\kappa x)$ behavior depending on its symmetry. The exact value of the mode parameter $\kappa$ is given by the requirement of the in-plane ($y,z$) field components continuity throughout $x=\pm d/2$ planes. For symmetric ($\cos$) mode profile, we obtain:
\begin{eqnarray}
\label{eqKappa}
\tan\left(\frac{\kappa^{(TM)}d}{2}\right)&=&-\frac{\epsilon_m \kappa^{(TM)}}
{\sqrt{{\rm k}^2\epsilon_{\|}^2(\epsilon_{\perp}-\epsilon_m)-\kappa^{(TM)^2}\epsilon_{\|}\epsilon_{\perp}}}
\\
\tan\left(\frac{\kappa^{(TE)}d}{2}\right)&=& \frac{\sqrt{{\rm k}^2 (\epsilon_{\|}-\epsilon_m)-\kappa^{(TE)^2}}}{\kappa^{(TE)}}
\nonumber
\end{eqnarray}

In the limit of $\epsilon_m\rightarrow-\infty$, these equations yield the values $\kappa_0=\pi(2j+1)/d$, used in the previous Section. As we previously noted, these values correspond to the well-known condition of zero mode magnitude at the waveguide boundary. In the limit of sufficiently large $|\epsilon_m|$ it is possible to find the correction to the above values of the mode parameter $\kappa$. Specifically, 
\begin{eqnarray}
\label{eqKapCorr}
\kappa^{(TM)}&\approx&
\kappa_0\left(1-\frac{2 {\rm k} \epsilon_{\|}}
{\kappa_0^2 d\sqrt{-\epsilon_m}}\right)
\\
\kappa^{(TE)}&\approx&\kappa_0\left(1-\frac{2}
{{\rm k} d\sqrt{-\epsilon_m}}\right)
\nonumber
\end{eqnarray}

\begin{figure}[htbp]
\centering
\includegraphics[width=8.3cm]{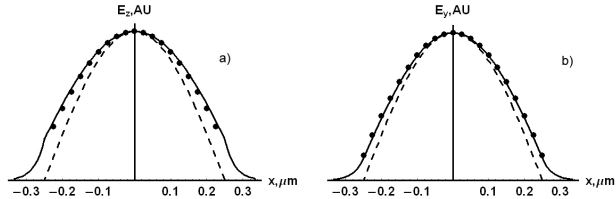}
\caption{The cross-section of the field in the planar waveguide with hollow $d=0.5\;\mu m$-thick core; dashed line corresponds to the case of $\epsilon_m=-\infty$ (perfect metal boundary); solid line corresponds to Ag boundaries for $\lambda=0.85\;\mu m$; dots correspond to $\kappa$ calculated using Eq.~(\ref{eqKapCorr}); TM (a) and TE (b) modes are shown}
\label{figModeStructure}
\end{figure}

As the mode parameter $\kappa$ plays a role of an inverse confinement length of the mode in $x$ direction, the negative $\kappa$ correction signifies the ``mode expansion'' into the waveguide wall region. Such a mode expansion is illustrated in Fig.~\ref{figModeStructure}.

The immediate effect of such a change in the mode structure is the change of the effective phase velocity, given by the refraction index: 
\begin{eqnarray}
\label{eqRefIndEff}
n^{(TM)}\approx\pm\sqrt{\epsilon_{\perp}\nu_0}\left(
1+\frac{2}{{\rm k}\; d\; \nu_0\sqrt{-\epsilon_m}}\right)
\\
n^{(TE)}\approx\sqrt{\epsilon_{\|}\nu_0}\left(
1+\frac{2\kappa_0^2}{{\rm k}^3 d\; \epsilon_{\|}\nu_0\sqrt{-\epsilon_m}}\right),
\nonumber
\end{eqnarray}
where $\nu_0=1-\kappa_0^2/(\epsilon_{\|}{\rm k}^2)$. As it has been described above, the sign of the refraction index for the $TM$ polarization has to be selected {\it positive} for $\epsilon_{\perp}>0; \; \nu>0$, and {\it negative} for $\epsilon_{\perp}<0; \; \nu<0$. 

Penetration of the mode into the waveguide wall region has another effect on the wave propagation. Namely, the finite value of the $\epsilon_m^{\prime\prime}$ introduces an additional (with respect to the core material) absorption into the system. As a result, the magnitude of a mode will exponentially decay as it propagates through the system. Such an attenuation can be related to the imaginary part of the effective refractive index through $E\propto \exp(-n^{\prime\prime}{\rm k} z)$. In the limit of small absorption in the metal ($\epsilon_m^{\prime\prime}/|\epsilon_m^{\prime}|\ll 1$) the ``waveguide-induced'' mode decay is described by: 
\begin{eqnarray}
\label{eqRefIndIm}
n^{(TM)^{\prime\prime}}\approx
\frac{1}{{\rm k} d}
\sqrt{\frac{\epsilon_{\perp}}{\nu_0|\epsilon_m|}}
\frac{\epsilon_m^{\prime\prime}}{|\epsilon_m^{\prime}|}
\\
n^{(TE) ^{\prime\prime}}\approx
\frac{\kappa_0^2}{{\rm k}^3 d \sqrt{\epsilon_{\|}\nu_0|\epsilon_m|}}
\frac{\epsilon_m^{\prime\prime}}{|\epsilon_m^{\prime}|}
\nonumber
\end{eqnarray}
Note that in agreement with causality principle\cite{podolskiyPRB,landauECM} the losses in the system are positive, regardless of the sign of the refractive index. 

Using Eq.~(\ref{eqRefIndIm}) we estimate that for wavelengths $\lambda\geq 850$~nm, the losses introduced by a silver waveguide walls are substantially small ($n^{\prime\prime}/n \lesssim 0.01$).

\section{Anisotropic nanoplasmonic composites}
We now consider the fabrication perspectives of the material with strong optical anisotropy required for NIM waveguide core region. A number of naturally occurring materials with the required properties exist at THz or far IR frequencies. Some examples include Bi and Sapphire\cite{podolskiyJMO}. Unfortunately, no known material exhibits anisotropy exceeding 30\% at optical or infrared spectral range. Here we propose to take advantage of a new class of nano-engineered media, known as meta-materials\cite{metamat}. In these composites, nanostructured particles are used as meta-atoms to achieve the desired EM properties. 

To realize the strong optical anisotropy we propose to use a combination of plasmonic or polar particles (providing the negative permittivity) and dielectric media (having $\epsilon>0$). If the characteristic size of inhomogeneities and their typical separation are much smaller than the wavelength of incident radiation, the EM response of the composite structure can be described in terms of the effective dielectric constant $\epsilon_{\rm eff}$\cite{landauECM}:
\begin{equation}
\label{eqEpsEffDef}
<D(r)>_{\alpha}=<\epsilon(r)_{\alpha,\beta}E(r)_{\beta}>=
\epsilon_{{\rm eff}_{\alpha,\beta}}<E(r)> _{\beta},
\end{equation} 
where the brackets ($<>$) denote the averaging over the microscopically-large (multi-particle), macroscopically small (subwavelength) spatial area, Greek indices denote Cartesian components, and summation over repeated indices is assumed. Since the size of a particle $a$ typically enters Maxwell equations in the combination ${\rm k}a$\cite{landauECM}, all size-related effects play a minor role in the considered here ``quasi-static'' averaging process. Therefore, we note that the composites proposed below are highly tolerant with respect to size variation. Also, since the effects described here are obtained in effective medium approximation, the desired response does not require any periodicity of the particle arrangement and only the {\it average concentration} has to be controlled during the fabrication step. 

Below we present two meta-material designs of the strongly anisotropic composite for optical and infrared spectrum ranges. 

\subsection{Layered system}

\begin{figure}[htbp]
\centering
\includegraphics[width=6cm]{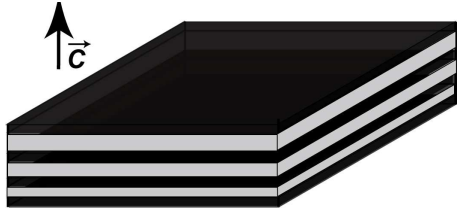}
\caption{The schematics of the layered structure described in the text}
\label{figLayersGeom}
\end{figure}

We first consider the permittivity of a stack of interlacing plasmonic (Ag,Au,Al,$\ldots$) or polar (SiC) ($\epsilon_{pl}<0$) and dielectric (Si,GaAs,$\ldots$) ($\epsilon_d>0$) layers. We assume that the layers are aligned in the waveguide ($y,z$) plane (see Fig.~\ref{figLayersGeom}). As noted above, the absolute thickness of the layers is not important (as long as it is subwavelength), and only the average concentration of plasmonic layers $N_{pl}$ plays the role. 

\begin{figure}[htbp]
\centering
\includegraphics[width=15cm]{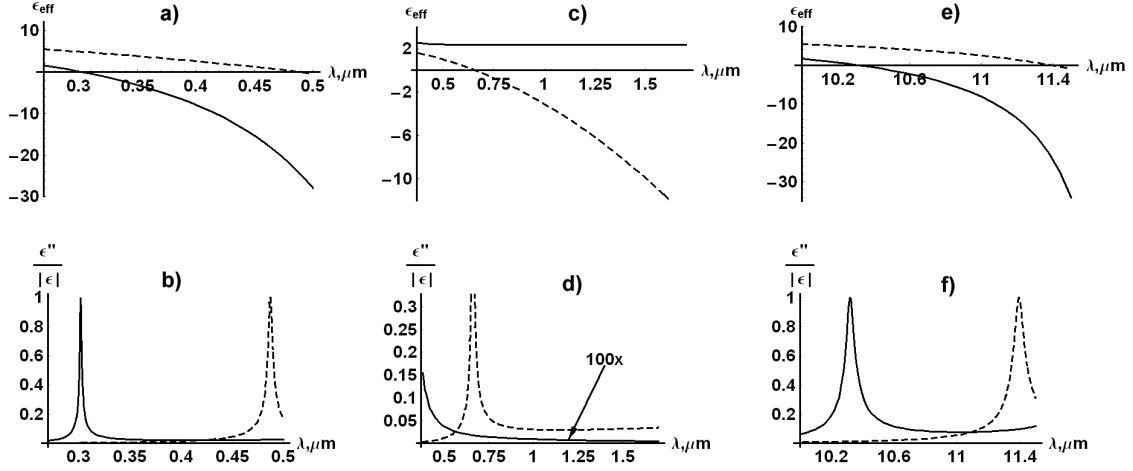}
\caption{Real part (a,c,e) and absorption (b,d,f) of effective $\epsilon_{\perp}$ (solid lines) and $\epsilon_{\|}$ (dashed lines) for layered systems; (a,b): $Ag-Si$ stack; $N_{\rm pl}=0.6$; (c,d): $Ag-SiO_2$ stack; $N_{\rm pl}=0.1$; Note the extremely small absorption of this system; (e,f) $SiC-Si$ stack; $N_{\rm pl}=0.1$. }
\label{figLayersDat}
\end{figure}

To compute $\epsilon_{\rm eff}$ we note that the $E_y, E_z$, and $\epsilon E_x$ have to be continuous throughout the system\cite{landauECM,pendryWires,shvetsWires}, leading to: 
\begin{eqnarray}
\label{eqEpsLyr}
\epsilon_{\|}=\epsilon_{{\rm eff}_{y,z}}=N_{pl}\; \epsilon_{pl}+
(1- N_{pl})\epsilon_d
\\
\epsilon_{\perp}=\epsilon_{{\rm eff}_x}=\frac{\epsilon_{pl}\epsilon_d}
{ (1- N_{pl}) \epsilon_{pl}+ N_{pl}\;\epsilon_d}
\nonumber
\end{eqnarray}

The effective permittivities for several layered composites are shown in Fig.~\ref{figLayersDat}. We note that while the strong anisotropy $\epsilon_{\|}\cdot\epsilon_{\perp}<0$ can be easily achieved in the layered system, the actual realizations of the materials with $\epsilon_{\|}>0, \epsilon_{\perp}<0$ required for the high-frequency NIM described here typically have substantial absorption\footnote{This particular realization of layered NIM structure for IR frequencies has been earlier proposed in Ref.~[\onlinecite{shvetsPRB}]}, and therefore have a limited range of applications\cite{podolskiyResolut,podolskiyResolut2}. 

On the contrary, the materials with $\epsilon_{\|}<0, \epsilon_{\perp}>0$ (achieved, for example by a repeated deposition of Ag-Si layers) form low-loss media. While this configuration has a positive refraction index, it may be potentially used to concentrate propagating modes in subwavelength areas\cite{podolskiyPRB,podolskiyInPress}. 

\subsection{Aligned wire structure}
\begin{figure}[htbp]
\centering
\includegraphics[width=15cm]{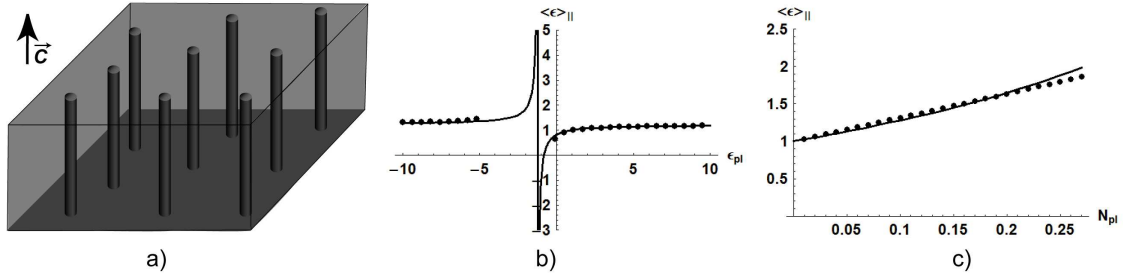}
\caption{(a) The schematics of the wired structure described in the text. (b-c) comparison of $<\epsilon>_{\|}$, calculated using Eq.~(\ref{eqEpsWir}) (solid line) and derived from numerical solution of Maxwell Eqs as described in the text (dots); dependence of $<\epsilon>_{\|}$ on dielectric constant of the inclusions for $N_{\rm pl}$ (b) and on concentration for $\epsilon_{pl}=-10$ (c) is shown.}
\label{figWiresGeom}
\end{figure}

The array of aligned $\epsilon_{pl}<0$ nanowires embedded in the dielectric ($\epsilon_d>0$) host, schematically shown in Fig.~\ref{figWiresGeom} is in some sense a counterpart of the layered system described above. In fact, the boundary conditions now require the continuity of the $E_x$ field, along with the solution of quasi-static equations in the $y-z$ plane. While in the general case the analytical solution of this problem is complicated, the case of small plasmonic material concentration is adequately described by the Maxwell-Garnett approximation\cite{landauECM,stroud,lakhtakiaJPhysD,podolskiyAniz}: 
\begin{eqnarray}
\label{eqEpsWir}
\epsilon_{\|}=\epsilon_{{\rm eff}_{y,z}}=
\frac{ N_{pl}\;\epsilon_{pl} E_{in}+(1- N_{pl})\epsilon_d E_0}
{ N_{pl} E_{in}+(1- N_{pl})E_0 }
\\
\epsilon_{\perp}=\epsilon_{{\rm eff}_{x}}= N_{pl} \epsilon_{pl}+
(1- N_{pl})\epsilon_d,
\nonumber
\end{eqnarray}
where $E_{in}=\frac{2\epsilon_d}{\epsilon_d+\epsilon_{pl}}E_0$ is the field inside the plasmonic inclusion and $E_0$ is the excitation field. 

To illustrate the validity of the MG approximation, we numerically solve the Maxwell equations in the planar geometry using the coupled-dipole approach (CDA), described in detail in Refs.~[\onlinecite{podolskiyOptExp,podolskiyAniz,podolskiyJOPA05}]. In these calculations the composite is represented by a large number of interacting point dipoles, and the resulting dipole moment distribution is related to the effective dielectric constant. Fig.~\ref{figWiresGeom} shows the excellent agreement between the numerical simulations and the analytical result [Eq.~(\ref{eqEpsWir})].

\begin{figure}[htbp]
\centering
\includegraphics[width=8.3cm]{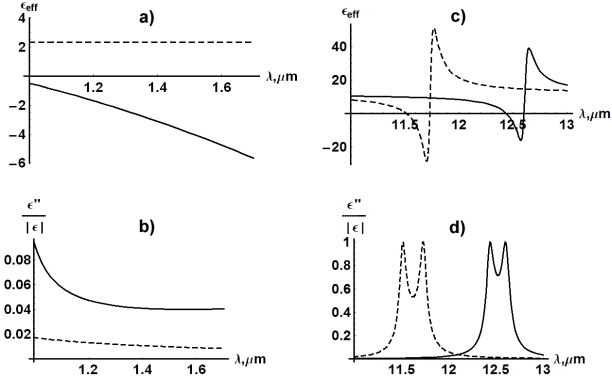}
\caption{ Real part (a,c) and absorption (b,d) of effective $\epsilon_{\perp}$ (solid lines) and $\epsilon_{\|}$ (dashed lines) for wired systems; (a,b): $Ag-SiO_2$ structure (note the relatively small absorption for the NIM regime); $N_{\rm pl}=0.05$; (c,d): $SiC-Si$ structure; $N_{\rm pl}=0.1$.}
\label{figWiresDat}
\end{figure}

The effective dielectric constants for some composite materials are presented in Fig.~\ref{figWiresDat}. Note that in contrast to the layered system described above, these wired composites have extremely low absorption in the near-IR-NIM regime –- in a way solving the major problem with the ``conventional'' design of optical LHMs. 

\section{Imaging Properties of non-magnetic optical NIMs}
To illustrate the imaging performance of the proposed system we calculate the propagation of a wavepacket formed by a double-slit source through the $5\mu m$-long planar layer of $5\%$ $Ag$, $95\%$ $SiO_2$ wire-based NIM-core described above [see Fig.~\ref{figWiresDat}(a,b)], embedded in the $Si$ waveguide. We select the thickness of the dielectric core to be $d=0.3\mu m$, and assume the excitation by the telecom-wavelength $\lambda=1.5 \mu m$. The Eqs.~(\ref{eqRefIndEff},\ref{eqEpsWir}) yield the following values of the refraction index: $n^{(+)}\approx 2.6$, $n^{(LHM)}\approx-2.6+0.05i$. 

To calculate the resulting field distribution we first represent the wavepacket at the $z=0$ plane as a linear combination of the waveguide modes\cite{podolskiyPRB,podolskiyJMO}. We then use the boundary conditions at the front and back interfaces of the NIM region to calculate the reflection and transmission of individual mode. The solutions of Maxwell equations are then represented as a sum of solutions for the individual modes. 

\begin{figure}[htbp]
\centering
\includegraphics[width=15cm]{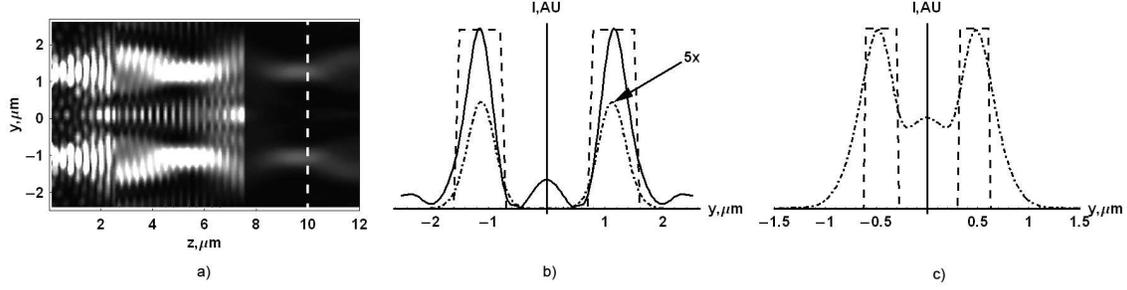}
\caption{Imaging by a planar NIM-based lens. $n>0$ region: $Si$-filled planar waveguide; $d=0.3 \mu m$; NIM region: planar waveguide with core material described in Fig.~\ref{figWiresDat}(a-b); (a) The intensity distribution in the system with absorption losses neglected; LHM region is beween $z=2.5\mu m$ and $z=7.5\mu m$; focal plane corresponds to $z=10 \mu m$ (white dashed line); slit size $w=0.75 \mu m$; (b) dashed line: emitted radiation; solid line: focal plane intensity distribution in system described in (a); dash-dotted line: same as solid line, but in the case of real (absorbing) NIM. (c) same as (b), but $w=0.3\mu m$ (corresponding to far-field resolution limit of the system)}
\label{figImaging}
\end{figure}

To better illustrate the imaging properties of the system and distinguish between the effects of negative refractive index and material absorption, we first neglect losses in the NIM core. The resulting intensity distribution in the system is shown in Fig.~\ref{figImaging}(a). The image formation in the focal plane ($z=10\mu m$) of the far-field planar NIM lens is clearly seen. 

In Fig.~\ref{figImaging}(b) we compare the imaging through the planar NIM lens with and without the material absorption and demonstrate that the presence of weak loss, although it reduces the magnitude of the signal, it does not destroy the far-field imaging. Similar to any far-field imaging system\cite{landauECM,podolskiyResolut,podolskiyResolut2}, the resolution $\Delta$ of the non-magnetic NIM structure presented here is limited by the internal wavelength: $\Delta\approx\lambda_{in}/2=\lambda/(2 n)\approx 0.3\mu m$ [see Fig.~\ref{figImaging}(c)]. 

\section{Conclusions}
In conclusion, we presented a non-magnetic non-periodic design of a system with negative index of refraction. We have further proposed several low-loss nanoplasmonic-based realizations of the proposed structure for optical and infrared frequencies. We have presented analytical description of the effective dielectric permittivity of strongly anisotropic nanostructured composites, and showed the excellent agreement of the developed theory with results of numerical solution of Maxwell equations. Finally, we have demonstrated the low-loss far-field planar NIM lens for $\lambda=1.5 \mu m$ with resolution $\Delta\approx 0.3 \mu m$. 

The authors would like to thank E.E.~Mishchenko and A.L.~Efros for fruitful discussions. The research was partially supported by NSF grants DMR-0134736, ECS-0400615 and Oregon State University. 









\begin{thebibliography}{99}
\bibitem{podolskiyResolut2} It can be shown \cite{podolskiyResolut,podolskiyResolutLong} that for losses $\frac{\{\epsilon,\mu\}^{\prime\prime}}{|\{\epsilon,\mu\}^{\prime}|}>0.3$ even the near-field resolution of the NIM-based system is smaller than that of conventional near-field optics.
\end{thebibliography}
\end{document}